\documentclass[twocolumn,twoside,preprintnumbers,amsmath,amssymb,pacs]{revtex4}
\usepackage{epsfig}
\usepackage{graphicx}% Include figure files
\usepackage{dcolumn}% Align table columns on decimal point
\usepackage{fancyhdr}

\usepackage{pslatex}

\begin{document}

\title{{\Large  Phenomenology of infrared finite gluon propagator and coupling constant  }}

\author{A. A. Natale}

\affiliation{ Instituto de F\'{\i}sica Te\'orica - UNESP - Rua Pamplona, 145 - 01405-900 - S\~ao Paulo - SP - Brazil }

%%%%%%%%%%%%\received{on September 14, 2006}

\begin{abstract}

We report on some recent solutions of the Dyson-Schwinger equations for the infrared
behavior of the gluon propagator and coupling constant, discussing their differences and proposing that these different behaviors can be tested through hadronic phenomenology. We discuss which kind of phenomenological tests can be applied to the gluon propagator and coupling constant, how sensitive they are to the infrared region of momenta and what specific solution is preferred by the experimental data. 



\end{abstract}

\maketitle

\thispagestyle{fancy}
\setcounter{page}{0}

\section{Introduction}

During the last years there has been much effort in trying to obtain the infrared
(IR) behavior of the Quantum Chromodynamics (QCD) Green's functions by means of 
theoretical and phenomenological studies \cite{r1,r2,r3,r4,r5,r6,r7,r8,r9,r10,r11} and by simulations of QCD on a lattice \cite{r12,r13,r14,r15}. The infrared behavior of the gluon and ghost propagators, as well as of the coupling constant are completely intertwined with the confinement problem. For example, if the gluon propagator would be as singular as $1/k^4$ when $k^2 \rightarrow 0$, it would indicate an interquark potential rising linearly with the separation.

In the late seventies, working in the Landau gauge and Euclidean space, Mandelstam obtained a solution of the Dyson-Schwinger equation (DSE) for the gluon propagator, in the case of pure gauge QCD, that behaved as $1/k^4$ \cite{r16}. A few years later Cornwall obtained a gauge invariant solution that behaved as $1/[k^2 + m^2(k^2)]$ \cite{r17}. In this last case, as $k^2 \rightarrow 0$, the function $m^2(k^2)$ was interpreted as a dynamical gluon mass with the limit $m^2(k^2\rightarrow 0) = m^2_g$. Both solutions reproduce the expected perturbative behavior naturally at large $k^2$. The $1/k^4$ result
was named as a ``confining solution", whereas the massive (or any infrared finite solution) became known as a ``confined solution". It is clear that for the confined solution a more sophisticated explanation of quark confinement must be at work, and for this reason the appealing $1/k^4$ solution became more popular, but not without some debate \cite{r18}.

Recently the coupled DSE equations for the gluon and ghost propagators were solved with
different approximations \cite{r2}, resulting in an infrared gluon propagator that
behaves roughly as $k^2/[k^4 + a^4]$, i.e. vanishes as $k^2 \rightarrow 0$. This kind of
behavior had also been predicted by Gribov \cite{r19} and Zwanziger \cite{r20}
in a different approach. The behavior of gluon and ghost propagators in Euclidean Yang-Mills theory quantized in the maximal Abelian gauge (MAG) were also studied considering
the effects arising from a treatment of Gribov copies in the MAG and those arising from
a dynamical mass originating in a dimension two gluon condensate \cite{r21}. In these studies the infrared gluon propagator depends on the so called Gribov parameter, whose variation seems to make an interpolation between the vanishing propagator and the massive one!
          
A great step ahead in this problem has also been provided by the QCD lattice
simulations in Landau gauge, which strongly support the existence of an infrared finite gluon propagator \cite{r12,r13,r14,r15}, where by finite we mean that the gluon propagator
may be zero or different from zero at $k^2 =0$. This is interesting enough, because indicates the appearance of a dynamical mass scale for the gluon, which imply in the existence of a non-trivial QCD infrared fixed point, i.e. the freezing of the coupling constant at the origin of momenta \cite{r8}. Unfortunately the lattice data is not precise enough to definitively settle the questions of Green's functions at the origin of momenta, but there are claims that the results can nicely accommodate a massive gluon solution
like the one proposed by Cornwall \cite{r13}. Notice that at first
glance it seems that these two possibilities are just a small detail, however they
imply in two different confinement scenarios. Cornwall indicated that a dynamically
generated gluon mass induces vortex solutions in the theory and these are responsible for
the quark confinement \cite{r17}. On the other hand the vanishing gluon propagator jointly
with an infrared enhanced ghost propagator may induce an effective linear confining potential between quarks \cite{r22}.
It is not surprising that DSE for the QCD propagators lead to different solutions as long as they are solved with different truncations and approximations, as, for instance, the
choice of the trilinear gluon vertex plays a crucial role in the solution. This means that
we should wait for still more improved DSE solutions or lattice simulations for the
gluon propagator. 

It is our purpose here to advocate that the hadronic phenomenology can
provide solid information about the infrared behavior for the 
gluon propagator and coupling constant.     
Which kind of phenomenological tests are possible to propose in order to study
the infrared behavior of the gluon propagator and coupling constant? In our opinion
there are two possibilities: ({\textsl{i}}) We can make use of QCD inspired models where it is
assumed ``a priori" that the gluon propagator and running coupling constant must have one
specific non-perturbative behavior, as, for example, happens in the Pomeron model of 
Landshoff and Nachtmann \cite{r23}; or ({\textsl{ii}}) we can consider perturbative QCD 
calculations introducing an effective propagator and coupling constant in the sense
of the ``Dynamical Perturbation Theory" (DPT) proposed by Pagels and Stokar \cite{r24}
many years ago. In the next section we will detail and discuss examples of these
two possibilities.

\section{Phenomenological tests for the infrared behavior of the gluon propagator and coupling constant}

\subsection{Testing the IR behavior in a QCD Pomeron model}

It is known that a connection between the theoretical predictions of the perturbative Pomeron described by the BFKL Pomeron \cite{r25} and the experimental 
data is far from being fully understood. Landshoff and Nachtmann proposed 
a simple Pomeron model where the gluon propagation is affected by the non-trivial QCD vacuum leading naturally to a correlation length 
(or gluon mass scale). The existence of this correlation length is the only feasible explanation to the fact that diffractive interactions mediated by the Pomeron (which is assumed to be composed by at least 
two gluons in a color singlet state exchanged in channel $t$ between hadrons) obey the additive quark rule. This model is quite suitable to test IR properties of 
coupling constant and gluon propagator obtained through DSE, because it always involves one integration over the product of gluons propagators and coupling constant ($g^2(k^2)\times
D(k^2)$).

As long as we deal with a QCD inspired model for the Pomeron we should be aware
that the model will be valid up to a certain scale or approximation. Comparing
the theoretical predictions of the Landshoff and Nachtmann (LN) Pomeron model with the 
experimental data we can be quite confident that the model is very reasonable to
compute diffractive scattering in the small transferred momentum limit (i.e. small
$t$ physics). The Pomeron, in the LN model, is described as the singlet channel 
exchange of two non-perturbative gluons between hadrons, and in the sequence we
discuss the case of elastic differential cross-section for proton-proton scattering.

In the LN model the elastic differential cross can be obtained from

\begin{equation}\label{dsigma}
  \frac{d\sigma}{dt}= \frac{|A(s,t)|^2}{16\pi s^2}
\end{equation}

\noindent
where the amplitude for elastic proton-proton scattering via
two-gluon exchange can be written as

\begin{equation}
A(s,t)= is8\alpha_{s}^2\left[T_{1} - T_{2}\right]
 \label{ampli}
\end{equation}

\noindent
with

\begin{equation}
T_{1}= \int_0^s\,d^{2}k \,
D\left(\frac{q}{2}+k\right)D\left(\frac{q}{2}-k\right)|G_{p}(q,0)|^{2} \label{t1}
\end{equation}

\begin{eqnarray}
T_{2}= \int_0^s 
&& d^{2}k \, D\left(\frac{q}{2}+k\right)D\left(\frac{q}{2}-k\right)G_{p}
\left(q,k-\frac{q}{2}\right)  \nonumber \\ 
&& \times\left[2G_{p}(q,0)-G_{p}\left(q,k-\frac{q}{2}\right)\right]
\label{t2}
\end{eqnarray}

\noindent
where $G_p (q,k)$ is a convolution of proton wave functions

\begin{equation}
G_p(q,k)= \int \, d^2pd\kappa \psi^\ast(\kappa,p)\psi(\kappa,p-k-\kappa q).
\label{gdp}
\end{equation}

In Eq.(\ref{t1}) $D(q^2)$ is the non-perturbative expression for the gluon propagator.
Note that the loop integral in Eq.(\ref{t1}) contains the 
product of two gluon propagators 
weighted by $G_p (q,k)$, which will be written in terms of proton form factors. This
is a general behavior, where the form factors will vary according to the type of
hadron involved in the scattering. Evidently this means that the integral will depend
on different scales contained in each form factor, which in the particular case of
the proton is the
proton mass, whereas for pion-proton, $\gamma - \rho$ scattering, or others, we have
different mass scales in the integration, in addition to the one that enters in the gluon
propagator and coupling constant. Another important feature of this
model is the kinematic structure, where 
one of the gluons carry most of the momentum. The other gluon just seems to enter
in the process with a very small momentum in order to form a color singlet,
but it is responsible for the most important part of the integration area \cite{r26}.
  
To obtain the cross section  for the elastic differential $p-p$ scattering we use
the gluon propagator in Landau gauge written as

\begin{equation} D_{\mu\nu}(q^2)= \left({\delta}_{\mu\nu}
-\frac{q_{\mu}q_{\nu}}{q^2}\right)D(q^2),
\label{landau}
\end{equation}

\noindent
where the expression for $D(q^2)$ obtained by Cornwall is given by

\begin{equation}\label{propcorn}
 D^{-1}(q^2) = \left[q^2 + M_g ^2(q^2)\right]bg^2\ln\left[\frac{q^2+ 4M^2_{g}}{\Lambda ^2}  \right],
\end{equation}
and the coupling constant equal to
\begin{equation} \alpha_{sC} (q^2)= \frac{4\pi}{b \ln\left[
(q^2 + 4M_g^2(q^2) )/\Lambda^2 \right]}, \label{acor} \end{equation}
where $M_g(q^2)$ is a dynamical gluon mass given by,
\begin{equation} M^2_g(q^2) =m_g^2 \left[
\frac{ \ln\left(\frac{q^2+4{m_g}^2}{\Lambda ^2}\right) }{\ln\left(\frac{4{m_g}^2}{\Lambda ^2}\right) }\right]^{- 12/11}
\label{mdyna} 
\end{equation}
$\Lambda$($\equiv\Lambda_{QCD}$) is the QCD scale parameter.

\noindent
We will also make use of the running coupling constant obtained by Fischer and
Alkofer \cite{r27} which is given by 

\begin{equation}\label{runalk}
  \alpha_{sA} (x)= \frac{\alpha_A(0)}{\ln (e + a_1x^{a_2} +b_1x^{b_2}) } ,
\end{equation}

\noindent
where $x=q^2$ and

\hskip 0.5in$  \alpha_A (0) = 2.972$,
\hskip 0.5in$  a_1= 5.292\;\mbox{GeV}^{-2a_2}$,

\hskip 0.5in$  a_2=2.324$,
\hskip 0.7in$  b_1= 0.034\;\mbox{GeV}^{-2b_2}$,

\hskip 0.5in$  b_2=3.169$.
and their respective propagator $ D(q^2) = Z(q^2)/q^2$,
where $Z(q^2)$, in Landau gauge, is fitted by

\begin{equation}\label{propgalk}
  Z(x)= \left(\frac{\alpha_{sA}(x)}{\alpha_{sA}(\mu)}\right)^{1+ 2\delta}R^2(x),
\end{equation}

\noindent
and

\begin{equation}\label{rpropg}
R(x)= \frac{cx^{\kappa}+ dx^{2\kappa}}{1 + cx^{\kappa}+
dx^{2\kappa} }
\end{equation}

\noindent
where the constants appearing in Eq.(\ref{propgalk}) and Eq.(\ref{rpropg}) are given by

\hskip 0.5in $\alpha_A (\mu^2) = 0.9676$,
\hskip 0.5in $\kappa= 0.5953$,

\hskip 0.5in $\delta=-9/44$,
\hskip 0.7in $c= 1.8934\;\mbox{GeV}^{-2\kappa}$,

\hskip 0.5in $d= 4.6944\;\mbox{GeV}^{-4\kappa}$.
  
All these expressions were obtained from DSE solutions. Eqs.(\ref{runalk}) and (\ref{propgalk})
were obtained in Landau gauge and correspond to a infrared vanishing 
gluon propagator. The gluon propagator comes with a definite mass scale once the renormalization procedure is defined and the coupling
constant fixed at one given scale. The Cornwall result was shown to be gauge invariant, and
many of the features of this propagator and coupling constant are discussed 
by Aguilar and Papavassiliou \cite{r28}, its mass scale depends on the ratio
$m_g/\Lambda$.

\begin{figure}[htbp]
\begin{center}
\includegraphics[width=8cm]{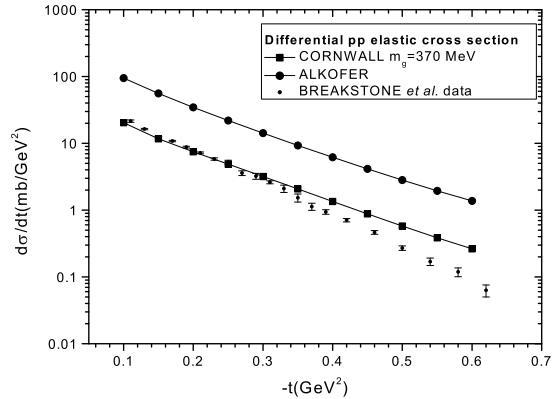}
\caption{Differential $pp$ elastic cross section at
$\sqrt(s)=53\, \mbox{GeV}$ computed within the
Landshoff--Nachtmann model for the Pomeron,
using different infrared couplings and gluon propagators
obtained from DSE solutions.} \label{transition}
\end{center}
\end{figure}

The details of the elastic differential $p-p$ scattering calculation can be found in Ref.\cite{r9}, and the
comparison of the result with the experimental data of
Breakstone {\it et al.} at $\sqrt{s}= 53$ GeV \cite{r29}is shown in Fig.(\ref{transition}).
Note that the agreement between theory and experimental data is good only
at small $t$, since at large $t$ values we have contributions from 3-gluon exchange.
We see in Fig.(\ref{transition}) that Cornwall's propagator with a dynamical 
gluon mass of \textsl{O}$(370)$MeV fits the data quite well, whereas the
Fischer and Alkofer one gives a quite large result for this cross section.
We obtained the same result when analysing total hadron-hadron cross section \cite{r9,r11}
and exclusive $\rho$ production in deep inelastic scattering \cite{r9}.  

The integration of Eq.(\ref{t1}) and Eq.(\ref{t2}) picks up an intermediate
region of momenta for the gluon propagators, i.e. above the region where
the propagator that vanishes at the origin is suppressed and the
coupling constant related to it is still large. As we shall see in the
next examples most of the problems with this solution comes from the
fact that the coupling constant given by Eq.(\ref{runalk}) is too large
at the origin and up to several hundred MeV, and the integrations that
we shall perform are not peaked at a momentum $k^2=0$, but are peaked in one region
where the product $g^2(k^2)D(k^2)$ appearing in Eq.(\ref{t1}) and (\ref{t2}) is not small. As discussed in Ref.\cite{r10} there
are many indications that the experimental data is better described
by a small coupling constant in the infrared.
 
\subsection{Testing the IR behavior in a perturbative QCD calculation - The $\gamma \rightarrow \pi^0$ transition }

Perturbative QCD calculations involve the exchange of gluons at large momenta. These
calculations do not provide very strong tests for the infrared behavior of the gluon
propagator, although a finite infrared gluon propagator gives a natural cutoff for the
many infrared divergent parton subprocess cross sections, which, in general, are
dependent on the cutoff choice. On the other hand they may be influenced by the
behavior of the infrared coupling constant that comes out from DSE solutions, which at intermediate momenta may differ appreciably from the perturbative behavior. It is
worth mentioning that several calculations in the literature make use of infrared
finite coupling constants. For example, they are natural in the so called Analytic Perturbation Theory (APT) \cite{r30} or they appear as an effective charge in
several perturbative QCD calculations \cite{r31}. In most of the cases these IR
finite coupling are fundamental to confront the theoretical
and experimental data for several physical quantities. As mentioned before, we shall use the non-perturbative
behavior of the coupling constant that comes out from DSE solutions in the
sense prescribed by DPT \cite{r24}.

The photon-to-pion transition form factor $F_{\gamma\pi}(Q^2)$ is measured in
single-tagged two-photon $e^+e^- \rightarrow e^+e^- \pi^0$ reactions. The
amplitude for this process has the factorized form

\begin{equation} \label{amppi}
F_{\gamma\pi}(Q^2) = \frac{4}{\sqrt{3}} \int^1_0 \, dx \, \phi_{\pi}(x,Q^2) T^H_{\gamma\pi}(x,Q^2),
\end{equation}

\noindent
where the hard scattering amplitude $ T^H_{\gamma\pi}(x,Q^2)$ is given by

\begin{equation} \label{hardgpi}
T^H_{\gamma\pi}(Q^2) = \frac{1}{(1-x)Q^2} [ 1 + {\cal O}(\alpha_s)].
\end{equation}

\noindent
Using an asymptotic form for the pion distribution amplitude $\phi_{\pi}=\sqrt{3} f_\pi x(1-x)$,
we obtain \cite{r10}

\begin{equation}\label{trans}
Q^2F_{\gamma\pi}(Q^2)=
2f_{\pi}\left(1-\frac{5}{3}\frac{\alpha_{V}(Q^\ast)}{\pi}\right)
\end{equation}

\noindent
where $Q^\ast = \exp^{-3/2} Q$ is the estimated Brodsky-Le\-pa\-ge-\-Mac\-ken\-zie scale
for the pion form factor in the scheme discussed in Ref.\cite{r32}.

\begin{figure}[htbp]
\begin{center}
\includegraphics[width=8cm]{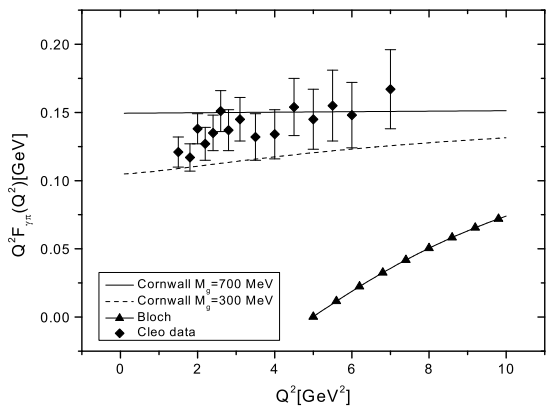}
\caption{The $\gamma \rightarrow \pi^{0}$ transition
form factor calculated with different expressions for the infrared
behavior of the running coupling constant.} \label{gamapi}
\end{center}
\end{figure}

In Fig.(\ref{gamapi}) we compare the photon to pion transition form factor
with CLEO data \cite{r33}. The curves were computed with different expressions for the
infrared behavior of the running coupling constant. We assumed
$f_{\pi} \simeq 93 \;\mbox{MeV}$ and  $\Lambda = 300$ MeV. 
We also made use of a running coupling constant determined by Bloch \cite{r34}, where
the absence of the Landau pole at $q^2=\Lambda^2$ is reminiscent of APT, which is given
by

\begin{eqnarray}\label{bloc}
\alpha_{sB}(q^2)&=& \alpha(l\Lambda_{QCD}^2) = \\ \nonumber
&&\frac{1}{c_{0} + l^2}\left[ c_{0}\alpha_0 +
\frac{4\pi}{\beta_0}\left( \frac{1}{\log(l)} -
\frac{1}{l-1}\right) l^2 \right]
\end{eqnarray}

\noindent
where $l=q^2/\Lambda^2_{QCD}$, $c_{0} = 15$, $\alpha_0 = 2.6$, and
$\beta_0 = 11 - \frac{2}{3}n_f$, where $n_f$ is the number of flavors. Eq.(\ref{bloc})
is also consistent with a propagator that vanishes at $k^2=0$. 
With the coupling constant of Eq.(\ref{bloc}) we obtain a fit for the photon-pion transition
form factor very far from the experimental data. The result obtained when we use Eq.(\ref{runalk})
is not shown and gives an even worse fit. The infrared value of the coupling
constant is so large in the case of the coupling constants given by Eqs.(\ref{runalk}) and (\ref{bloc}),
that we are not sure that the perturbative result can be trusted even at such large momentum scale. The momentum
scale appearing in Fig.(\ref{gamapi}) is above the GeV scale, where the DSE solutions
for the gluon propagators already assume their perturbative $1/k^2$ behavior. 
%%%%%%%%%%%%%%%%%%%%%%%%
%%%%%%%%%%%%%%%%%%%%%%%
The values of the infrared coupling constants related to the class of DSE solutions consistent with
a vanishing gluon propagator  are much stronger
than most of the phenomenological estimates of the frozen $\alpha_s(0)$ value
that we quoted in Ref.\cite{r10} ($\alpha_s (0) \approx 0.7 \pm 0.3$),
and are at the origin of the strange lower curve of Fig.(\ref{gamapi}). 
The data is only compatible with
Eq.(\ref{acor}), which has a smoother increase towards the infrared region. Perhaps
this behavior is actually indicating that the transition to the infrared should be
a soft one.  

Note that in Fig.(\ref{gamapi}) the curves obtained with
Cornwall's coupling constant do not show large variation in the full range of
uncertainty of the dynamical gluon mass. It is interesting that
its behavior is quite stable in this case as well as for the pion form factor
studied in Ref.\cite{r10}. If we had large variations of the infrared coupling
constant with the gluon mass scale we could hardly propose any reliable
phenomenological test for its freezing value. We stress that the results are
obtained for a perturbative scale of momenta and it seems that we have to choose
between two possibilities: ({\it i}) This process cannot be predicted by perturbative
QCD up to a scale of several GeV, or ({\it ii}) some of the DSE solutions are predicting a
too large value of the coupling constant in the infrared and the approximations
made to determine these solutions are too crude.

We have also computed the perturbative pion form factor ($F_\pi (q^2)$), and the same happens in that case, i.e. only 
the solutions for the gluon propagator and coupling constant determined by
Cornwall, when plugged into the perturbative QCD expression for $F_\pi$, produce the match between the theory and the experimental
data. The details of the pion-photon transition
form factor calculation can be seen in Ref.\cite{r9} and the pion form factor
calculation is described in Ref.\cite{r10}. 
 
\subsection{Testing the IR behavior in a perturbative QCD calculation - Total
hadronic cross sections within the parton model}

Total hadronic and jet cross sections can be calculated in a straightforward way
within the parton model (see Ref.\cite{r35} for a review);
The cross section
for producing jets with $p_{T}>p_{T_{min}}$ through the dominant process $gg\to gg$ is given by
\begin{eqnarray}
\sigma_{jet}(s) = &&\int_{p^2_{T_{_{min}}}} dp^2_T \,
 \frac{d\hat{\sigma}_{gg}}{dp^2_T} \times \nonumber \\
&&\int_{x_1 x_2 > 4p^2_T /s} dx_1 dx_2 \, g(x_{1},Q^2) \, g(x_{2},Q^2) ,
\label{eq1}
\end{eqnarray}
where $g(x,Q^2)$ is the gluon flux and $p^2_{T_{min}}$ is the
momentum above which we can use the perturbative calculation of the subprocess differential
cross section ${d\hat{\sigma}_{gg}}/{dp^2_T}$. This momentum scale is usually assumed to be
larger than $1$ GeV$^2$. At high energies, the first order perturbative result for the cross section
$\hat{\sigma}_{gg}$ can be written in terms of the variable $\hat{s}$  as \cite{r36}
\begin{eqnarray}
\hat{\sigma}_{gg}(\hat{s})  = \frac{9\pi \alpha_0^2}{m_0^2} \, \theta (\hat{s} -m_0^2) ,
\label{eq3}
\end{eqnarray}
where $m_{0}$ is a particle production threshold and $\alpha_0$ is an effective value of the running coupling constant.
This cross section causes a rapid increase in $\sigma_{jet}(s)$ if a eikonalization
procedure is not used in the calculation. Assuming the simple ansatz for
the gluon flux
\begin{eqnarray}
g(x) \sim \frac{(1-x)^5}{x^J},
\label{eq4}
\end{eqnarray}
a straightforward calculation yields
\begin{eqnarray}
\sigma_{jet}(s) \sim \frac{9\pi \alpha_0^2}{m_0^2} \left(\frac{s}{m_0^2}\right)^\epsilon ,
\label{eq5}
\end{eqnarray}
where $\epsilon \equiv J-1 > 0$. In this calculation we have neglected some factors in the right hand side of the final
expression. Nevertheless, in the limit of large enough $s$ this expression reproduces the expected
asymptotic energy dependence of $\sigma_{jet}(s)$. Moreover, with specific values for $m_{0}$ and
$\alpha_{0}$, it is possible to show
that at $\sqrt{s}\sim 630$ GeV, this jet cross section is of order of the $pp$ and $\bar{p}p$ total cross sections
\cite{r36}. It is also clear that Eq. (\ref{eq5}) does violate unitarity.

The simple derivation of the cross section behavior that we
have seen above is not too useful due to the following reasons: a) the $m_0^2$ and the $\alpha_0^2$ terms in
Eq. (\ref{eq5}) are totally {\it ad hoc} and b) Eq. (\ref{eq5}) violates unitarity. Unitarity
is recovered with the eikonalization of the model. However the procedure still keeps the {\it ad hoc}
constants $m_0^2$ and $\alpha_0^2$ as parameters that can be obtained only with the data fitting. In Ref. \cite{r37} the elementary gluon-gluon cross section was calculated within the dynamical
perturbation theory scheme (DPT) \cite{r24}, where the effective gluon propagator and coupling constant
enters into the calculation. With this procedure we eliminate the freedom existent in the
previous calculation, due to the choice of  $m_0^2$ and $\alpha_0^2$, changing $m_0^2$ by
$m_g^2$ and still have a parameter less in the calculation because the running coupling
now is also a function of  $m_g^2$.

In the eikonal representation the total cross section is
given by
\begin{eqnarray}
\sigma_{tot}(s) = 4\pi \int_{_{0}}^{^{\infty}} \!\!  db\, b\, [1-e^{-\chi_{_{I}}(b,s)}\cos \chi_{_{R}}(b,s)],
\label{eq21}
\end{eqnarray}
where $s$ is the square of the total center-of-mass energy, $b$ is the impact parameter, and
$\chi(b,s)=\chi_{_{R}}(b,s)+i\chi_{_{I}}(b,s)$ is a complex eikonal function. In the QCD eikonal model with a
dynamical gluon mass, henceforth referred to as DGM model, we write the even eikonal as the sum of
gluon-gluon, quark-gluon, and quark-quark contributions:
\begin{eqnarray}
\chi^{+}(b,s) &=& \chi_{qq} (b,s) +\chi_{qg} (b,s) + \chi_{gg} (b,s) \nonumber \\
&=& i[\sigma_{qq}(s) W(b;\mu_{qq}) + \sigma_{qg}(s) W(b;\mu_{qg}) \nonumber \\
&+& \sigma_{gg}(s) W(b;\mu_{gg})] ,
\label{eq22}
\end{eqnarray}
where $\chi_{pp}^{\bar{p}p}(b,s) = \chi^{+} (b,s) \pm \chi^{-} (b,s)$. Here $W(b;\mu)$ is the overlap
function in the impact parameter space and $\sigma_{ij}(s)$ are the elementary
subprocess cross sections of colliding quarks and gluons ($i,j=q,g$). The overlap function is associated with
the Fourier transform of a dipole form factor,
\begin{eqnarray}
W(b;\mu) = \frac{\mu^2}{96\pi}\, (\mu b)^3 \, K_{3}(\mu b),
\label{eq23}
\end{eqnarray}
where $K_{3}(x)$ is the modified Bessel function of second kind. The odd eikonal $\chi^{-}(b,s)$, that
accounts for the difference between $pp$ and $\bar{p}p$ channels, is parametrized as
\begin{eqnarray}
\chi^{-} (b,s) = C^{-}\, \Sigma \, \frac{m_{g}}{\sqrt{s}} \, e^{i\pi /4}\, 
W(b;\mu^{-}),
\label{eq24}
\end{eqnarray}
where $m_{g}$ is the dynamical gluon mass and the parameters $C^{-}$ and $\mu^{-}$ are constants to be
fitted. The eikonal functions $\chi_{qq} (b,s)$ and
$\chi_{qg} (b,s)$, needed to describe the low-energy forward data, are parametrized with terms
dictated by the Regge phenomenology:
\begin{eqnarray}
\chi_{qq}(b,s) = i \, \Sigma \, A \,
\frac{m_{g}}{\sqrt{s}} \, W(b;\mu_{qq}),
\label{eq26}
\end{eqnarray}
\begin{eqnarray}
\chi_{qg}(b,s) = i \, \Sigma \left[ A^{\prime} + B^{\prime} \ln \left( \frac{s}{m_{g}^{2}} \right) \right] \,
W(b;\sqrt{\mu_{qq} \, \mu_{gg}}),
\label{eq27}
\end{eqnarray}
where $A$, $A^{\prime}$, $B^{\prime}$, $\mu_{qq}$ and $\mu_{gg}$ are fitting parameters. 

The innovation in our approach is that the subprocesses cross sections that appear in
Eq.(\ref{eq22}) are computed within DPT, where the gluon propagator and coupling constants
are the ones obtained from DSE solutions. For example, $\sigma_{gg}(s)$ is computed
with the help of the Cornwall propagator and coupling constant, where, for
simplicity, we neglected
the momentum dependence of the running mass in the denominator of the gluon propagator,
obtaining 
\begin{eqnarray}
\hat{\sigma}^{D\!PT}(\hat{s}) &=& \frac{3\pi \bar{\alpha}_{s}^{2}}{\hat{s}} 
\Biggl[ \frac{12\hat{s}^{4} - 55 m_{g}^{2} \hat{s}^{3} + 12 m_{g}^{4} 
\hat{s}^{2} + 66  m_{g}^{6} \hat{s} - 8 m_{g}^{8}}{4 m_{g}^{2} 
\hat{s} [\hat{s} - m_{g}^{2}]^{2}} \nonumber \\ 
&& - 3  \ln \left( \frac{\hat{s} - 3m_{g}^{2}}{m_{g}^{2}}\right)\Biggr]  .
\label{h6}
\end{eqnarray}

\begin{figure}[htbp]
\begin{center}
\includegraphics[width=8cm]{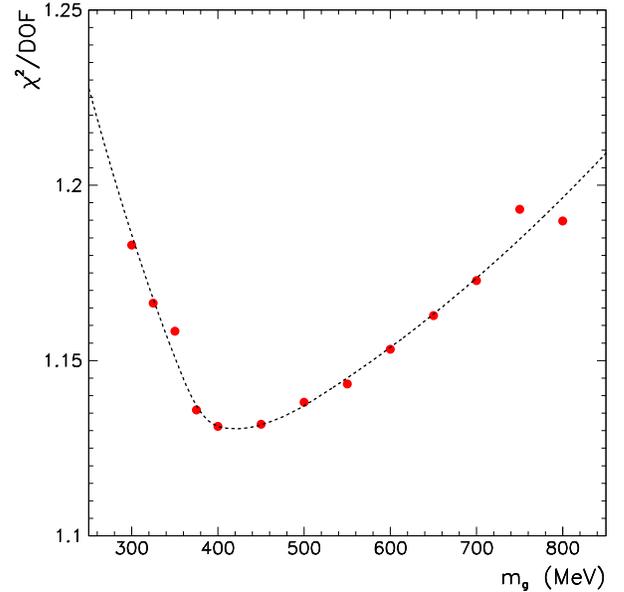}
\caption{The $\chi^{2}/DOF$ as a function of dynamical gluon mass $m_{g}$.} 
\label{difdad}
\end{center}
\end{figure}

\begin{figure}[htbp]
\begin{center}
\includegraphics[width=8cm]{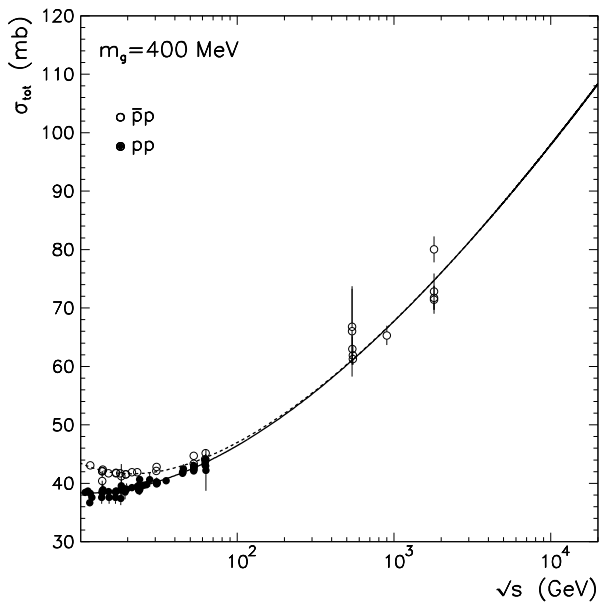}
\caption{Total cross section for $pp$ (solid curve) and $\bar{p}p$ (dashed curve) scattering.} 
\label{difdad2}
\end{center}
\end{figure}

The full detail of the calculation can be found in Ref.\cite{r37}. We fitted
the $pp$ and $p\bar{p}$ scattering data keeping $m_g$ as a free parameter. Our global fit results indicate a minimum value just about $m_{g}\approx 400$ MeV.
These results are shown in Fig.(\ref{difdad}), where a general dashed curve is added to guide the eye. Roughly, taking a
5\% variation on the minimal $\chi^{2}/DOF$ value indicated by the general curve, it is possible to estimate a
dynamical gluon mass $m_{g}\approx 400^{+350}_{-100}$ MeV. This value obtained
through data fitting is totally compatible with the ones found by Cornwall and subsequent determinations \cite{r11,r17}. The
fits for the total cross sections ($\sigma_{tot}$) can be seen in Fig.(\ref{difdad2}) in the case of
$m_{g} = 400$ MeV. 

It must be noticed that in the cross sections calculations we do have another integration
over the gluon propagators and coupling constant expressions weighted by the parton structure functions. In this case we also test the infrared behavior of the gluon propagator because
a large part of high energy behavior of the total hadronic cross section is due to
soft gluons \cite{r35}, which means small $\hat{s}$ values in Eq.(\ref{h6}) corresponding to the IR contribution of the gluon propagator. An analysis similar to the $pp$ total
cross section calculation shown is this subsection has also been performed in the
case of $\gamma - p$ and $\gamma - \gamma$ scattering with the same dynamical gluon
mass \cite{r38}.

\section{Conclusions}

We discussed DSE solutions for the gluon propagator. These solutions can be divided in two
possibilities according to their behavior as $k^2\rightarrow 0$, one vanishes at the origin and the other has a value different from zero and consistent with a dynamically generated
gluon mass. We should not be surprised with the existence of different solutions because
they are originated from different approximations or truncations of the DSE. As one example
we could recall that for many years an IR DSE solution for the gluon propagator behaving
like $1/k^4$ was very popular, but recent lattice QCD simulations basically discarded such possibility.

Lattice QCD simulations have already proportioned a great improvement in our knowledgement
of infrared Green's functions. There are strong indications that the gluon propagator
is finite and possibly the same may happens for the coupling constant. This result is 
important because it may give some information about the confinement mechanism.
Unfortunately the lattice result is still not precise enough near the origin of momenta in
order to discriminate between the different gluon propagator solutions.

In this work we are advocating that hadronic phenomenology can distinguish between
the different IR behaviors of DSE. No matter how we deal with 
QCD inspired models, like the LN Pomeron model, or just perturbative QCD calculations
improved by the use of an effective gluon propagator or coupling constant, we see that
Cornwall's propagator is selected by the experimental data. In all the examples
discussed in the previous section, we see that the phenomenological information is 
non-trivial, in the sense that it results from the calculation of physical quantities where the gluon propagator or product of propagators are integrated
weighted by different functions (involving different mass scales), and all quantities
show agreement with the experimental data for gluon masses that are in the same
range of masses predicted by Cornwall several years ago. It is hard to believe that
such coincidence is a fortuitous one. These values for the gluon mass lead to a frozen coupling constant at the origin of momenta whose value is not larger than \textsl{O}(1). Certainly an infrared 
finite coupling constant is welcome for hadronic phenomenology \cite{r10,r39}. Maybe future lattice calculations will shed some light on this problem, indicating which DSE approximation is reasonable or not, which confinement scenario is more probable to be
at work, and which kind of improved phenomenological calculation can be done with
such infrared finite gluon propagator and coupling constant.

\begin{acknowledgments}
I thank A. C. Aguilar and E. G. S. Luna for discussions and collaboration on
the topics presented here. I also thank the IRQCD 06 organizers for the
hospitality.
This research was partially supported by the Conselho
Nacional de Desenvolvimento Cient\'{\i}fico e Tecnol\'ogico-CNPq under contract 301002/2004-5.
\end{acknowledgments}

\end{document}